\documentclass[sigconf]{acmart}

\AtBeginDocument{%
  \providecommand\BibTeX{{%
    \normalfont B\kern-0.5em{\scshape i\kern-0.25em b}\kern-0.8em\TeX}}}

\usepackage{bbm}
\usepackage{xcolor}
\def\BibTeX{{\rm B\kern-.05em{\sc i\kern-.025em b}\kern-.08em
		T\kern-.1667em\lower.7ex\hbox{E}\kern-.125emX}}

\copyrightyear{2021} 
\acmYear{2021} 
\setcopyright{acmcopyright}\acmConference[ICAIF'21]{2nd ACM International Conference on AI in Finance}{November 3--5, 2021}{Virtual Event, USA}
\acmBooktitle{2nd ACM International Conference on AI in Finance (ICAIF'21), November 3--5, 2021, Virtual Event, USA}
\acmPrice{15.00}
\acmDOI{10.1145/3490354.3494377}
\acmISBN{978-1-4503-9148-1/21/11}

\begin{document}

\title{Deep Risk Model: A Deep Learning Solution for Mining Latent Risk Factors to Improve Covariance Matrix Estimation}

\author{Hengxu Lin}
\authornote{The first two authors have equal contribution.}
\authornote{This work was done when the first author was an intern at Microsoft Research Asia.}
\affiliation{%
  \institution{Columbia Business School}
  \city{New York}
  \country{United States}
}
\email{helin23@gsb.columbia.edu}

\author{Dong Zhou}
\authornotemark[1]
\affiliation{%
  \institution{Microsoft Research}
  \streetaddress{No.5 Danling Street}
  \city{Beijing}
  \country{China}
  \postcode{100080}
}
\email{Zhou.Dong@microsoft.com}

\author{Weiqing Liu}
\affiliation{%
  \institution{Microsoft Research}
  \streetaddress{No.5 Danling Street}
  \city{Beijing}
  \country{China}
  \postcode{100080}
}
\email{Weiqing.Liu@microsoft.com}

\author{Jiang Bian}
\affiliation{%
  \institution{Microsoft Research}
  \streetaddress{No.5 Danling Street}
  \city{Beijing}
  \country{China}
  \postcode{100080}
}
\email{Jiang.Bian@microsoft.com}


\begin{abstract}
Modeling and managing portfolio risk is perhaps the most important step to achieve growing and preserving investment performance. Within the modern portfolio construction framework that built on Markowitz's theory, the covariance matrix of stock returns is a required input to calculate portfolio risk. Traditional approaches to estimate the covariance matrix are based on human-designed risk factors, which often require tremendous time and effort to design better risk factors to improve the covariance estimation. In this work, we formulate the quest of mining risk factors as a learning problem and propose a deep learning solution to effectively ``design'' risk factors with neural networks. The learning objective is also carefully set to ensure the learned risk factors are effective in explaining the variance of stock returns as well as having desired orthogonality and stability. Our experiments on the stock market data demonstrate the effectiveness of the proposed solution: our method can obtain $1.9\%$ higher explained variance measured by $R^2$ and also reduce the risk of a global minimum variance portfolio. The incremental analysis further supports our design of both the architecture and the learning objective.
\end{abstract}



\maketitle

\section{Introduction} \label{sec:intro}
Modern portfolio theory has witnessed the increasing significance of risk management in stock investment in past decades. Successful risk management recognizes the equity market risk that the stocks have exposure. Thereby enables investors to adjust their strategies according to the estimation of returns and arising uncertainty while constructing their portfolio. The most widely adopted framework to select the best stock portfolio is built on Markowitz's mean-variance theory~\cite{markowitz1952portfolio}. Within this framework, the optimal portfolio is obtained by solving a constrained utility-maximization problem, with the utility defined as the expected portfolio return minus a cost of portfolio risk. To precisely determine the risk of any portfolio, having a good estimation of the covariance matrix of stock returns is the fundamental question.

Nevertheless, the empirical covariance matrix can be extremely ill-conditioned as the dimension of stocks grows and even becomes rank deficient such that cannot be used for portfolio optimization. To facilitate the estimation of the covariance matrix under such high-dimension setting, \cite{ledoit2004honey,ledoit2012nonlinear} proposed to shrink the off-diagonal elements in the empirical covariance matrix towards zero, based on the assumption that the target covariance matrix is sparse. However, stock returns are often highly related to each other and thus the covariance matrix of stock returns cannot sparse. A more realistic assumption is that the high-dimensional stock returns are driven by a set of low-dimensional \emph{risk factors}~\cite{sheikh1996barra}. For example, it is often observed that stocks with similar market capitalization tend to have similar market behavior, thus the market capitalization can be such a risk factor, which is also widely known as the \emph{size}~\cite{fama1992} factor. By introducing multiple risk factors, the estimation of a high-dimensional covariance matrix of stock returns can be decomposed to the estimation of a low-dimensional covariance matrix of risk factor returns\footnote{Factor returns are the cross-sectional regression coefficients. We will use these two terms interchangeably in the rest of the paper.}. Apparently, a good estimation of the covariance matrix requires a good design of the risk factors.

Risk factors have been carefully studied by investment professionals. Fundamental risk factors are proposed first since they are highly interpretable on an ongoing basis. Capital assets pricing model (CAPM)~\cite{capm} first formulated the systematic risk factor known as \emph{beta}. Multiple factors are further included into this fundamental risk factor family: size and value~\cite{fama1992}, momentum~\cite{carhart1997persistence} and non-linear size (BARRA)~\cite{sheikh1996barra}, to name a few. 
Despite their success, the design of these risk factors was extremely slow: it took almost half a century from CAPM to BARRA, with only roughly ten new factors added in the evolution. This is because the discovery of these risk factors usually relies on tremendous human efforts to verify hypotheses based on the long-term observation. Statistical risk factors~\cite{alexander2001market,avellaneda2010statistical} are thereby proposed to address such limitation, which applies matrix decomposition algorithms like principal component analysis (PCA) or Factor Analysis~\cite{harman1976modern} to obtain the latent risk factors. However, such decomposition methods are linear, thus hard to discover non-linear factors like non-linear size as aforementioned. Besides, unlike fundamental risk factors that have been tested in explaining forward returns, statistical risk factors are simply the principle components of history returns and often struggle in out-of-sample generalization. Therefore, a more efficient and effective risk factor mining solution is still desired for improving the estimation of covariance matrix of stock returns.

In this work, we propose a deep learning solution, \emph{Deep Risk Model}, to facilitate the design of risk factors. From the perspective of machine learning, the task of designing risk factors could be effectively solved given a well-designed architecture to fully utilize the available information and an appropriate learning objective to match the desired properties of risk factors. First of all, the architecture should not only be compatible with the transformations used in human-designed risk factors, but also have the capacity to represent more complex transformation. To this end, we design a hybrid neural network with Gated Recurrent Units (GRU)~\cite{chung2014empirical} to model the temporal transformation (e.g., the standard deviation of historical stock returns), and a Graph Attention Networks (GAT)~\cite{velivckovic2017graph} layer to represent the cross-sectional transformation (e.g., the excess return over the sector average). Furthermore, we design a learning objective to pursue the desired properties of the learned risk factors on three dimensions: the ability of explaining the variance of stock returns, orthogonality and stability.

To demonstrate the effectiveness of the proposed method, we further conduct extensive experiments on the stock market data. Specifically, we first compare the variance explanation ability of risk factors from the proposed method with the ones from the fundamental risk model and statistical risk model. Experiment results show our method can consistently outperform the other two baselines with $1.9\%$ higher $R^2$. We further use the learned risk factors to estimate the covariance with which we construct a Global Minimum Variance (GMV) portfolio. We find the proposed deep risk model’s factors are more reliable and achieve the lowest volatility for the GMV portfolio. Furthermore, we analyze the characteristics of the learned factors and receive excellent accomplishment of controlling multi-collinearity and stability. Ablation studies on our model also manifest the necessity and superiority of our architecture and learning objective design.

The main contributions are summarized as follows:
\begin{itemize}
    \item To the best of our knowledge, we are the first to formulate the quest of risk factor mining as a supervised learning task.
    \item We propose \emph{Deep Risk Model} as an effective and adaptive solution for learning risk factors with a comprehensive design of both the architecture and the learning objective.
    \item We conduct extensive experiments with real-world stock market data and compare with existing state-of-the-art baselines. Experiment results demonstrate the effectiveness of the proposed method.
\end{itemize}

\section{Related Works} \label{sec:related}
\paragraph{Covariance Matrix Estimation}
Large scale covariance matrix estimation is fundamental in multivariate analysis and ubiquitous in financial economics panel data. The main challenge of this field remains on the singular matrix arising from an insufficient sample size compared to the large sample dimension. Related solutions mainly focus on two aspects: rank-based estimation and factor-based estimation. Rank-based estimation relies on the sparsity assumption of the matrix: a majority of the matrix elements is nearly zero, thus various thresholds~\cite{bickel2008covariance,rigollet2012estimation,lam2009sparsistency} could be designed to control these elements and the parameters to be estimated are reduced significantly. Factor-based estimation is widely adopted in the scenario where the sparsity assumption is not applicable when variables are highly correlated with each other (e.g. stocks). By introducing factors and the factor covariance matrix to the decomposition, the residual covariance could be expressed as a conditional sparse matrix: conditional to common factors, the covariance matrix of remaining components are sparse. The factor model is more reasonable and commonly used in economics forecasting for its capacity to perform well in out-of-sample test~\cite{fan2011high,feng2020taming,fan2018large}.

\paragraph{Matrix decomposition}
Matrix Decomposition or Matrix Factorization, which is fundamental in linear algebra~\cite{trefethen1997numerical,fan2016overview}, factorizing the matrix into a product of matrices. Matrix decomposition algorithms are designed for specific purpose. LU decomposition, QR decomposition, Cholesky decomposition, rank factorization are used to solve systems of linear equations. Eigendecomposition, or spectral decomposition is used to attain eigenvectors and eigenvalues. Jordan decomposition, Schur decomposition, scale invariant decomposition~\cite{uhlmann2018generalized}, singular value decomposition (SVD)~\cite{van1976generalizing} are further implementations derived from eigendecomposition. In addition, principal component analysis (PCA)~\cite{wold1987principal} is an application of SVD. 

\paragraph{Representation Learning}
Representation Learning, or Feature Learning, is a technique used to extract effective representations needed for specific downstream tasks (e.g. classification, regression, feature detection) from raw features. Unsupervised and supervised representation learning has been well studied for decades: matrix factorization as discussed above, unsupervised dictionary learning, principal components analysis, independent components analysis~\cite{vasilescu2005multilinear}, clustering techniques like K-Means, linear discriminant analysis~\cite{balakrishnama1998linear}. Recently, the success of deep learning brings multiple supervised representation learning techniques into prosperity. Neural networks have demonstrated their distinguished competency and accomplished state-of-the-art in representation learning even for various kinds of tasks: vision computing~\cite{caron2018deep,asano2020self}, natural language processing~\cite{mikolov2013efficient,sutskever2014sequence,devlin2019bert}.

\section{Preliminaries} \label{sec:foundation}
The main interest of this section is to give an overview of the modern multi-factor model theory used in \emph{factor-based risk models}, and to manifest the necessity of adopting an advanced solution to discover and capture intrinsic and adaptive risk factors. 
\subsection{Factor-Based Risk Models}
Driven by the assumption that the intrinsic dependency of stock returns could be explained by some common latent factors, a large volume of academic and industrial research discusses and studies the design of factors. Mathematically, the multi-factor model is a multi-variable linear model with risk factors as the independent variables and stock return as the dependent variable. Let $\mathrm{y_{i,t}}$ denote the observable return of stock $i$ at time $t$, the multi-factor model takes the form of

\begin{equation}\label{eq:reg}
    \mathrm{y}_{i,t}=\mathbf{f}_{i}^\intercal\mathbf{b}_t+\mathrm{u}_{i,t}
\end{equation}
where $\mathbf{f}_{i} \in \mathbb{R}^{K}$ represents a vector of $K$ risk factors for the $i$-th stock, $\mathrm{u}_{i,t}$ is the idiosyncratic error term, and $\mathbf{b_t} \in \mathbb{R}^K$ is the vector of regression coefficients, which is also called factor returns. 

Moreover, considering forward returns of a specific stock over some horizon $H$, we can further decompose the stock returns into factor returns by
\begin{equation}
\begin{aligned}
\begin{bmatrix}
\mathrm{y}_{i,t} \\
\mathrm{y}_{i,t+1} \\
... \\
\mathrm{y}_{i,t+H-1} \\
\end{bmatrix} = 
\begin{bmatrix}
\mathbf{f}_{i}^\intercal\mathbf{b}_{t} + \mathrm{u}_{i,t} \\
\mathbf{f}_{i}^\intercal\mathbf{b}_{t+1} + \mathrm{u}_{i,t+1} \\
... \\
\mathbf{f}_{i}^\intercal\mathbf{b}_{t+H-1} + \mathrm{u}_{i,t+H-1} \\
\end{bmatrix},
\end{aligned}
\end{equation}
or equivalently in its matrix form
\begin{equation} \label{eq:ts-reg}
    \mathbf{y}_i = \mathbf{B} \mathbf{f}_i + \mathbf{u}_i,
\end{equation}
where $\mathbf{B} \in \mathbb{R}^{H \times K}$ contains the factor returns from time $t$ to $t+H-1$. Without loss of generality, we assume that $\mathbf{E}[\bar{\mathrm{y}}_i]=0$ and $\mathbf{E}[\bar{\mathrm{u}}_i]=0$, hereby, the covariance $\Sigma_{ij}$ of any two stocks $i$, $j$ can be derived as:
\begin{equation} 
    \Sigma_{ij} = \mathbf{f}_i \Sigma_{\mathbf{B}}\mathbf{f}^\intercal_{j}+\sigma_i \rho_{ij} \sigma_j
\end{equation}
where $\Sigma_\mathbf{B}$ is the covariance of the factor return matrix $\mathbf{B}$, $\sigma_i$ is standard deviation of the idiosyncratic errors $\mathbf{u}_i$, and $\rho_{ij}$ denotes the correlation between $\mathbf{u}_i$ and $\mathbf{u}_j$. A common assumption is that the idiosyncratic errors are independent (otherwise they can be explained by certain risk factors) and thus we have $\rho_{ij}=1$ if $i=j$ and $\rho_{ij}=0$ otherwise. 

Additionally, let $\mathbf{F}:=[\mathbf{f}_1,...,\mathbf{f}_N]^\intercal \in \mathbb{R}^{N \times K}$ denote the risk factor matrix for all $N$ stocks, the entire covariance matrix of stock returns can be derived as
\begin{equation} \label{eq:cov}
    \Sigma = \mathbf{F}\Sigma_{\mathbf{B}}\mathbf{F}^\intercal + \Delta,
\end{equation}
where $\Delta$ is a diagonal matrix with $\Delta_{ii} = \sigma_i^2$. Thereby we have decomposed the covariance of stock returns $\Sigma \in \mathbb{R}^{N \times N}$ into the covariance of factor returns $\Sigma_\mathbf{B} \in \mathbb{R}^{K\times K}$ which has lower dimension\footnote{As we cannot access all forward returns $\mathrm{y}_{i,t+1}$ at time $t$, a common practice is using historical stock returns to estimate $\Sigma_{\mathbf{B}}$.}. Apparently, the design of risk factors $\mathbf{F}$ is crucial to give accurate covariance estimation as both $\mathbf{b}$ and $\mathbf{u}$ are determined when $\mathbf{F}$ is specified in Equation~\ref{eq:reg}.

\subsection{Risk Factor Design}
The most popular risk factors are designed by human experts~\cite{sheikh1996barra}, which have been grounded in academic research and historically recognized to explanations of stock return: Size (measures a company’s scale of market capital), Value (measures a stock’s intrinsic value), Beta (explanation to market index return), Momentum (uses price trend to forecast future return), to name a few.
Despite the achievement from the human-designed risk factors, it actually took a long time for human specialists to draw these factors based on the long-term performance. Besides, evidence also exemplifies that these factors have become invalid during unpredictable market movements. Statistical risk factors are thereby proposed to address such limitations of human-designed factors, which usually applies principal component analysis (PCA) on historical stock returns to obtain the first $K$ components as the risk factors. Although efficient, such a method often has poor out-of-sample performance in risk forecasting as the obtained risk factors simply overfit the history returns but not explain the forward-looking returns. The limitations of both these two methods stimulate us to design a more efficient and effective solution for the design of risk factors. 

\section{Deep Risk Model} 

From the perspective of machine learning, the design of risk factors can be seen as the learning of a transformation function $\emph{g}$, which maps raw data $\mathbf{X}$ to risk factors $\mathbf{F}$ by $\mathbf{F}=\emph{g}(\mathbf{X})$. From this perspective, human-designed factors are equivalent to specifying the transformation $\emph{g}$ as explicit formulas. In this work, we consider adopting a deep neural network to learn a parameterized transformation $\emph{g}_\theta$ in a data-driven approach. Apparently, a well-designed neural network architecture should not only be compatible with human-designed transformation formulas but also have the capacity to capture more complex transformations. Therefore, we design a hybrid neural network to represent both temporal transformation and cross-section transformation, besides the neural network can also ensure capture the complex non-linear transformations. Moreover, we also prepare our learning objective carefully to ensure the ability of learned risk factors to explain the variance of stock returns, while satisfying the requirements of both orthogonality and stability. 

\subsection{Architecture Design} \label{sec:arch}
\begin{figure}[ht]
  \centering
  \includegraphics[width=0.9\columnwidth]{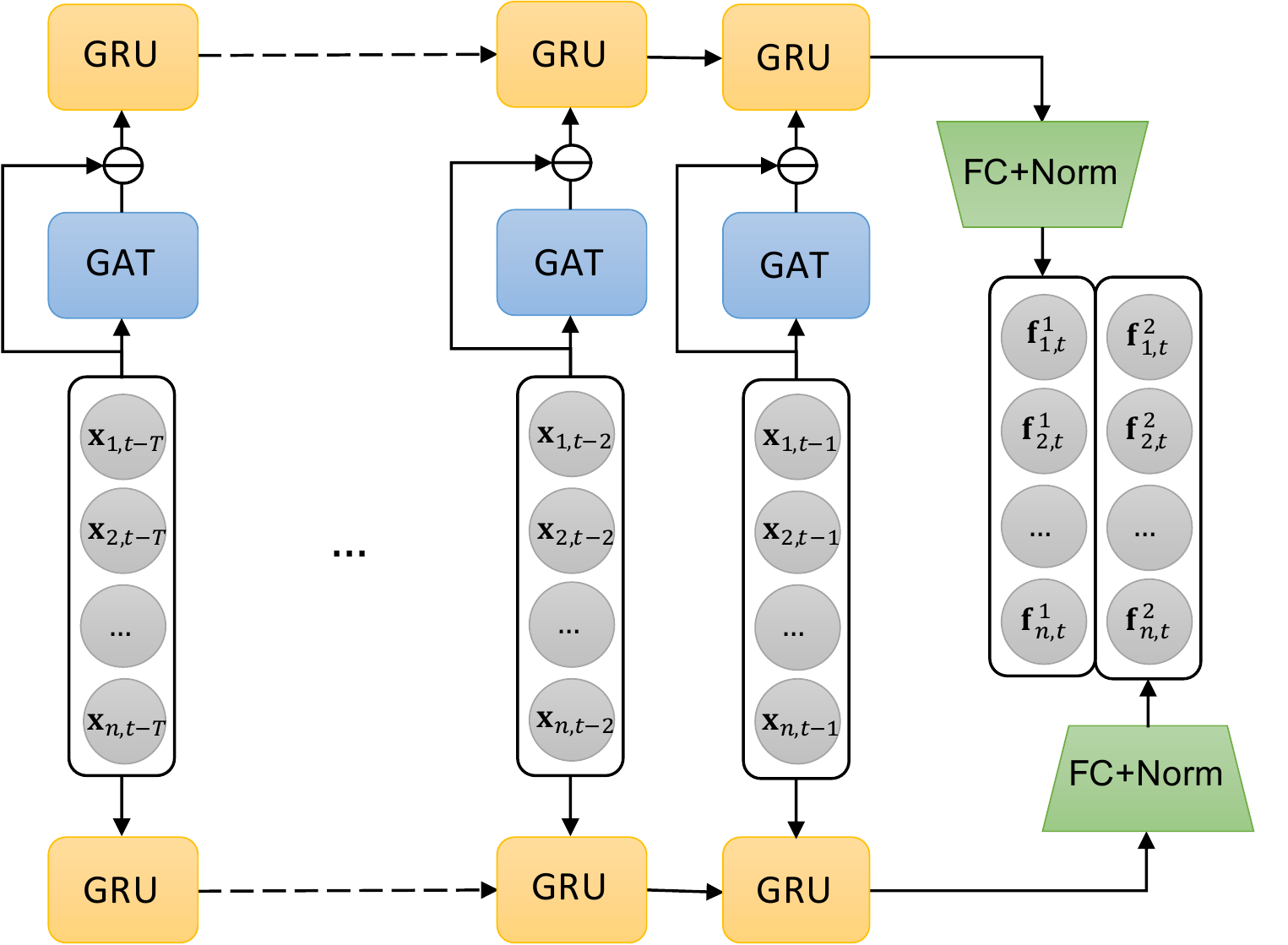}
  \caption{The proposed architecture for risk factors mining consists of two branches of Gated Recurrent Units (GRU) modules to learn temporal transformations. One of the two branches has a Graph Attention Network (GAT) layer to learn cross-sectional transformation.}
  \label{fig:framework}
\end{figure}

\subsubsection{Temporal Transformation}
In order to learn the temporal transformation, e.g., the standard deviation of historical returns, to a full extent from the panel data, we employ Gated Recurrent Units (GRU)~\cite{chung2014empirical} as the backbone model. GRU has been widely used in natural language processing and receives remarkable achievements in modeling sequential information. Denote the input feature as $\mathbf{x}_{i,t} \in \mathbb{R}^P$ for stock $i$ at time $t$ with $P$ as the number of features. The general idea of GRU is to recurrently project the input time series into the hidden representation distribution space. At each time step, the GRU learns the hidden representation $\mathbf{h}_{i,t}$ by jointly leveraging current input $\mathbf{x}_{i,t}$ and previous hidden representation $\mathbf{h}_{i,t-1}$ in a recursive manner:
\begin{equation}
    \mathbf{h}_{i,t} = \mathrm{GRU}(\mathbf{x}_{i,t}, \mathbf{h}_{i,t-1}).
\end{equation}
Besides, in order to adaptively select the most important information from different time steps, we also use the attention mechanism~\cite{feng2018enhancing,HengxuKDD2021} to aggregate all hidden representations.

\subsubsection{Cross-Sectional Transformation}

The aforementioned temporal transformation simply treats each stock independently, which, however, is unable to represent the transformation that involves relations among different stocks. For example, a common transformation in human-designed factors is subtracting stock returns from the average return of stocks that belongs to the same sector to get the residual momentum factor~\cite{blitz2011residual}. Such a transformation requires the model to be aware of the relations of individual stocks in addition to the temporal transformation. To fulfill this goal, we introduce another branch of GRU with an additional Graph Attention Networks (GAT)~\cite{velivckovic2017graph} layer to model the cross-sectional transformation as illustrated in Future~\ref{fig:framework}. The GAT layer not only can represent the traditional stock relations like sector, but can also dynamically capture latent relations with learnable attention weights.

Mathematically, given the raw features $\mathbf{X} \in \mathbb{R}^{N \times P}$ for $N$ stocks at time $t$, the GAT layer first projects the input features into another space with a shared linear transformation $\mathbf{W} \in \mathbb{R}^{P \times P}$, and then performs self-attention $a: \mathbb{R}^P \times \mathbb{R}^P \to \mathbb{R}$ on the nodes (i.e., stocks) to compute the attention coefficients:
\begin{equation}\label{eq:att}
    e_{ij} = \mathrm{LeakyReLU}\big( a(\mathbf{W}\mathbf{x}_{i}, \mathbf{W}\mathbf{x}_{j}) \big).
\end{equation}
The attention coefficients can indicate the similarity of stock $i$ and stock $j$. A prior graph structure, e.g., sector or business relation, can also be injected into GAT by performing masked attention~\cite{velivckovic2017graph}.

The second step in GAT is aggregating the information from stocks with high attention coefficients, which can be accomplished by normalzing the attention coefficients by the softmax function:

\begin{equation}\label{eq:softmax-att}
    \alpha_{ij} = \mathrm{softmax}_j(e_{ij}) = \frac{\mathrm{exp}(e_{ij})}{\sum_{k \in \mathcal{N}_i} \mathrm{exp}(e_{ik})},
\end{equation}
where $\mathcal{N}_i$ contains all neighborhoods of the $i$-th stock.
The normalized attention coefficients $\alpha_{ij}$ can thereby be used to compute a linear combination of the features by
\begin{equation}\label{eq:out-gat}
    \mathbf{\tilde{x}}_i = \mathrm{LeakyReLU}\Big(\sum_{j\in\mathcal{N}_i} \alpha_{ij} \mathbf{W}\mathbf{x}_{i} \Big).
\end{equation}
A shared GAT layer will be applied at all time stamps to get the aggregated information $\mathbf{\tilde{x}}_{i,t}$. As we are interested in learning the relative information compared to the group of the related stocks for cross-sectional transformation, we will subtract the aggregated information from neighborhoods from the current feature and pass the residuals to the GRU module:
\begin{equation}
    \mathbf{\tilde{h}}_{i,t} = \mathrm{GRU}(\mathbf{x}_{i,t} - \mathbf{\tilde{x}}_{i,t}, \mathbf{\tilde{h}}_{i,t-1}),
\end{equation}
where we use another GRU to further accomplish the temporal transformation from the residual features. The output hidden representations from both branches will be project into the risk factors by two separate linear projections $\mathbf{Q} \in \mathbb{R}^{P \times K_1}$ and $\mathbf{\tilde{Q}} \in \mathbb{R}^{P \times (K - K_1)}$:
\begin{equation} \label{eq:norm}
    \mathbf{f}_{i,t} = \big[\mathrm{Norm} ( \mathbf{Q} \mathbf{h}_{i,t}) \| \mathrm{Norm} ( \mathbf{\tilde{Q}} \mathbf{\tilde{h}}_{i,t}) \big],
\end{equation}
where $\|$ represents the concatenation operation, and $\mathrm{Norm}$ is a normalization function that transforms the learned risk factors to have capitalization weighted zero mean and equal-weighted unit standard deviation~\cite{balakrishnama1998linear}.

\subsection{Loss Design}
Now we ponder the design of loss function (i.e. the learning objective) to fulfill three desired properties of the mined risk factors. 

\subsubsection{Explained Variance} \label{sec:loss-goal}
The most important metric to evaluate the quality of a given set of risk factors is the proportion of the variance of stock returns predicted with the risk factors. In statistics, the coefficient of determination, $R^2$, is used to measure the ratio of explained variance. Let $\mathbf{F}_{\cdot t}=\emph{g}_{\theta}(\mathbf{X}_{\cdot t})$ denote the risk factors produced by our model at time $t$ for all stocks, then $R^2_{\cdot t}$ can be calculated as 
\begin{equation}\label{eq:r2}
R^2_{\cdot t} = 1 - \frac{\big\| \mathbf{y}_{\cdot t} - \mathbf{F}_{\cdot t} (\mathbf{F}_{\cdot t}^\intercal \mathbf{F}_{\cdot t})^{-1}\mathbf{F}_{\cdot t}^\intercal \mathbf{y}_{\cdot t} \big\|_2^2}{\| \mathbf{y}_{\cdot t} \|_2^2},
\end{equation}
where $||\cdot||_2^2$ represents the $\ell^2$ distance. Thence, we can use the below empirical loss function to optimize the model towards higher $R^2$:

\begin{equation}\label{eq:r2_loss}
    \max \frac{1}{T}\sum_{t=1}^{T} R^2_{\cdot t} \triangleq \min_\theta \frac{1}{T}\sum_{t=1}^{T} \frac{\big\| \mathbf{y}_{\cdot t} - \mathbf{F}_{\cdot t} (\mathbf{F}_{\cdot t}^\intercal \mathbf{F}_{\cdot t})^{-1}\mathbf{F}_{\cdot t}^\intercal \mathbf{y}_{\cdot t} \big\|_2^2}{\| \mathbf{y}_{\cdot t} \|_2^2}.
\end{equation}

\subsubsection{Multi-collinearity Regularization}
Directly training the model towards minimizing the objective defined in Equation~\ref{eq:r2_loss} often ends up with highly correlated risk factors, which will make the regression coefficients in Equation~\ref{eq:reg} unstable and thus harm the estimation of the covariance matrix. A practical statistical ratio used for diagnosing multi-collinearity is variance inflation factor (VIF)~\cite{belsley2005regression}, which quantifies the extent of correlation between $i$-th variate $\mathbf{F}_{i\cdot}$ in all time and the remaining covariates:
\begin{equation}\label{eq:vif}
    \mathrm{VIF}_{i\cdot} = \frac{1}{1-R_{i\cdot}^2} 
\end{equation}
where $R_{i\cdot}^2$ is the coefficient of determination for the regression of the $i$-th risk factors $\mathbf{F}_{i\cdot}$ on the other covariates. A straight forward idea to reduce the multi-collinearity is using $\sum_{i=1}^K \mathrm{VIF}_{i\cdot}$ as the regularization term, which however, requires $K$ separate regressions to obtain all VIF ratios and is time consuming. 

In this work, we introduce a computationally efficient alternative to obtain $\sum_{i=1}^K \mathrm{VIF}_{i\cdot}$. In the remainder of this section, we use $\mathbf{F}_{-i\cdot}$ to denote all the remaining factors excluding the $i$-th factor. Let $\mathbf{F}_{-i\cdot}$ be the independent variables and $\mathbf{F}_{i\cdot}$ as the dependent variable and through linear regression we can obtain the regression coefficients $\boldsymbol{\beta}_{-i} \in \mathbb{R}^{K-1}$. Then we have

\begin{equation}
\begin{aligned}
    \sum_{i=1}^K \mathrm{VIF}_{i\cdot} &=
    \sum_{i=1}^K (1-R_{i\cdot}^2)^{-1} \\&=
    \sum_{i=1}^K \mathbf{F}_{i\cdot}^\intercal\mathbf{F}_{i\cdot}[(\mathbf{F}_{i\cdot}-\mathbf{F}_{-i\cdot}\boldsymbol{\beta}_{-i})^\intercal(\mathbf{F}_{i\cdot}-\mathbf{F}_{-i\cdot}\boldsymbol{\beta}_{-i})]^{-1} \\&=
    \sum_{i=1}^K \mathbf{F}_{i\cdot}^\intercal\mathbf{F}_{i\cdot}[\mathbf{F}_{i\cdot}^\intercal\mathbf{F}_{i\cdot}-2\boldsymbol{\beta}_{-i}^\intercal\mathbf{F}_{-i\cdot}^\intercal\mathbf{F}_{i\cdot}+\boldsymbol{\beta}_{-i}^\intercal\mathbf{F}_{-i\cdot}^\intercal\mathbf{F}_{-i\cdot}\boldsymbol{\beta}_{-i}]^{-1}
\end{aligned}
\end{equation}
Since the factors are normalized to have unit deviation in Equation~\ref{eq:norm}, we substitute $\mathbf{F}_{i\cdot}^\intercal\mathbf{F}_{i\cdot}=N$. And from standardized regression characteristics we also have $\mathbf{F}_{-i\cdot}^\intercal\mathbf{F}_{-i\cdot}\boldsymbol{\beta}_{-i} = \mathbf{F}_{-i\cdot}^\intercal\mathbf{F}_{i\cdot}$, then:

\begin{equation}
\begin{aligned}
    \sum_{i=1}^K \mathrm{VIF}_{i\cdot} &= 
    N\sum_{i=1}^K \big(\mathbf{F}_{i\cdot}^\intercal\mathbf{F}_{i\cdot}-\boldsymbol{\beta}_{-i}{-i\cdot}^\intercal\mathbf{F}_{-i\cdot}^\intercal\mathbf{F}_{i\cdot}\big)^{-1} \\&= 
    N\sum_{i=1}^K \big(\mathbf{F}_{i\cdot}^\intercal\mathbf{F}_{i\cdot}- \mathbf{F}_{i\cdot}^\intercal \mathbf{F}_{-i\cdot} (\mathbf{F}_{-i\cdot}^\intercal \mathbf{F}_{-i\cdot})^{-1} \mathbf{F}_{-i\cdot}^\intercal\mathbf{F}_{i\cdot}\big)^{-1} \\&=
    N\sum_{i=1}^K \big(\mathbf{F}^\intercal\mathbf{F}\big)_{i\cdot,i\cdot}^{-1} = N * \mathrm{tr}\big(\big(\mathbf{F}^\intercal\mathbf{F}\big)^{-1}\big).
\end{aligned}
\end{equation}
The second to last equation is supported by Schur complement without loss of generality. Compared with regressing each variate $i$ on the remaining covariates to obtain $\sum_{i=1}^K \mathrm{VIF}_{i\cdot}$, the trace of matrix $(\mathbf{F}^\intercal\mathbf{F})^{-1}$ is more feasible and efficient. Consequently, we can regularize multi-collinearity by minimizing $\mathrm{tr}\big(\big(\mathbf{F}^\intercal\mathbf{F}\big)^{-1}\big)$.

\subsubsection{Multi-Task Learning}
Another desired property of the risk factors is their stability, i.e., the value of risk factors should be continuous along the time. This is an important property as in Equation~\ref{eq:ts-reg} and Equation~\ref{eq:cov} we assume the risk factors are constant in the estimation period. To ensure the stability of the risk factors, we design a multi-task learning objective that requires the learned risk factors $\mathbf{F}_{\cdot t}$ can be used to explain the variance of stock returns in multiple forward periods:
\begin{equation}\label{eq:multi-task}
    \min_{\theta} \frac{1}{H}\sum_{h=1}^{H} \frac{\big\|\mathbf{y}_{\cdot,t+h} - \mathbf{F}_{\cdot t}(\mathbf{F}^\intercal_{\cdot t}\mathbf{F}_{\cdot t})^{-1}\mathbf{F}^\intercal_{\cdot t}\mathbf{y}_{\cdot t}\big\|_2^2}{\|\mathbf{y}_{\cdot,t+h}\|_2^2},
\end{equation}
where $H$ is the number of forward periods.

Finally, we put together all the loss terms from the above discussion and propose the following learning objective:
\begin{equation}\label{eq:loss}
    \min_{\theta} \frac{1}{T} \sum_{t=1}^T \Big [\frac{1}{H}\sum_{h=1}^{H} \frac{\big\|\mathbf{y_{\cdot, t+h}} - \mathbf{F}_{\cdot t}(\mathbf{F}^\intercal_{\cdot t}\mathbf{F}_{\cdot t})^{-1}\mathbf{F}^\intercal_{\cdot t} \mathbf{y_{\cdot t}}\big\|_2^2}{\|\mathbf{y_{\cdot, t+h}}\|_2^2} + \lambda  \mathrm{tr} \big( (\mathbf{F}^\intercal_{\cdot t}\mathbf{F}_{\cdot t})^{-1} \big) \Big]
\end{equation}
where $\lambda$ is a hyper-parameter that controls the strength for the multi-collinearity regularization.

\section{Experiments}

\subsection{Experiment Settings}

\subsubsection{Data} \label{sec:data}
The proposed framework can accept all common structured data like price, financial data, or even text, as input features to train the risk model. In this work, we use $10$ common style factors in~\cite{orr2012supplementary} as the input features, including \emph{Size, Beta, Momentum, Residual Volatility, Non-linear Size, Book-to-Price, Liquidity, Earnings Yield, Growth} and \emph{Leverage}. All these style factors will also serve as risk factors in the compared fundamental risk model to make sure our method have no advantage of using external information in the experiments. In fact, this is a more challenging experiment setting as these risk factors are summarized from decades of research efforts and our model must be able to learn better risk factors to beat human experts. These factors are implemented for China stock market and the studied universe covers all listed stocks. We will follow the temporal order to split data into training (2009/04/30$\sim$2014/12/31), validation (2015/01/01$\sim$2016/12/31) and testing (2017/01/01$\sim$2020/02/10).

\subsubsection{Model Implementation} \label{sec:impl}
We implement the architecture designed in Section~\ref{sec:arch} with PyTorch. Both the two GRU modules in the designed network have $2$ hidden layers and $32$ hidden units. We let the number of learned risk factors equal to $K/2$ for both branches. The GAT module has the same hidden size as the input size, which is $10$. We also add attention dropout in GAT with dropout probability set to $0.5$. We pack all stocks in the same day as a single batch to ensure GAT can well capture stock relations and leverage gradient accumulation with the update frequency set to $64$ to ensure the model see as many diverse samples as possible. We use Adam~\cite{kingma2014adam} as the optimizer with a fixed learning rate set as $0.0002$. To further ensure the stability of the predictions, we also add exponential smoothing to the predictions with a smoothing factor set to $0.99$. Unless otherwise specified, the regularization parameter $\lambda$ is set to $0.01$. The number of tasks in Equation~\ref{eq:loss} is set to $20$, i.e., we encourage the learned risk factors can be used in the next $20$ days. The model is trained on a single TITAN Xp GPU and it takes $6.5$ hours to finish the model training for learning $10$ risk factors.

\subsection{Evaluation Protocol}
\subsubsection{Compared Methods} We compare the following methods for designing risk factors:
\begin{itemize}
    \item \textbf{Fundamental Risk Model (FRM)}: \textbf{FRM} directly uses the $10$ style factors described in Section~\ref{sec:data} as risk factors. \textbf{FRM} shows the best performance of the risk factors designed by human experts.
    \item \textbf{Statistical Risk Model (SRM)}: \textbf{SRM} takes the first $10$ components with the largest eigenvalues by applying PCA on stock returns in the last $252$ trading days. \textbf{SRM} shows the best performance of the traditional data-driven approach for learning latent risk factors.
    \item \textbf{Deep Risk Model (DRM)}: \textbf{DRM} is the proposed method in this work, which learns $10$ risk factors directly from the data described in Section~\ref{sec:data}.
\end{itemize}
As both \textbf{SRM} and \textbf{DRM} can learn any number of latent risk factors, we also add the results of \textbf{SRM (2x)} and \textbf{DRM (2x)} which will learn $20$ risk factors.

\subsubsection{Covariance Estimation}\label{sec:cov}
After the acquisition of the risk factors with different risk models, we will combine the risk factors with the country factor and $29$ industry factors for cross-sectional regression with Equation~\ref{eq:reg}. As the country factor introduces an exact collinearity into this equation, we follow~\cite{Menchero2011TheBU} to introduce an extra constraint to force the coefficients of industry factors to have zero sum. After obtaining the factor returns, we then follow Equation~\ref{eq:cov} to derive the covariance matrix. To further ensure the estimated covariance more responsive to the market, we estimate the correlation and variance with an exponential halflife of $240$ and $60$ days respectively and we also apply Volatility Regime Adjustment ~\cite{Menchero2011TheBU} with an exponential halflife of $20$ days to alleviate the forecasting bias.

\subsubsection{Evaluation Metrics} \label{sec:metrics}
To measure the risk forecasting performances of different risk models, we consider the below metrics:
\begin{itemize}
    \item $R^2$: $R^2$ measures the proportion of the explained variance of stock returns when using different sets of risk factors. The calculation of $R^2$ is defined as Equation~\ref{eq:r2} and we will report the average $R^2$ in the test periods. A good risk model should have higher $R^2$.
    \item \textbf{GMV}: We further show the performance of the estimated covariance matrix with different risk factors by constructing a Global Minimum Variance (GMV) portfolio and reporting its annualized volatility. Given a covariance matrix $\Sigma$, the GMV portfolio can be solved as $\frac{\boldsymbol{\Sigma}^{-1} \mathrm{1}}{\mathrm{1}^\intercal \boldsymbol{\Sigma}^{-1} \mathrm{1}}$ where $\mathrm{1}$ is a vector of ones. A good risk model should ensure the GMV portfolio has lower out-of-sample volatility.
    \item \textbf{GMV+}: This has the same objective as the GMV portfolio, except an extract no short-selling constraint is added. We use CVXPY~\cite{diamond2016cvxpy} to solve the optimized portfolio.
\end{itemize}

To further show the characteristics of the learned risk factors, we will also report the following statistics:
\begin{itemize}
    \item \textbf{Mean |t|}: This is the average absolute t-statistic of the regression coefficients in Equation~\ref{eq:reg}. A higher absolute t-statistic means the factor is more significant.
    \item \textbf{Pct |t|>2}: As we have multiple t-statistics across the whole studied periods, we also report the percent of days when the corresponding risk factor is statistical significant ($|t|>2$).
    \item \textbf{VIF}: This is the variance inflation factor (VIF) defined in Equation~\ref{eq:vif}, which shows whether the current factor is collinear with the remaining factors. A lower VIF is always preferred.
    \item \textbf{Auto Corr.}: Auto correlation measures the stability of the risk factor. We will report the auto correlation of each factor with 1 day's lagging. A higher correlation is preferred.
\end{itemize}

\subsection{Main Results}

\subsubsection{Risk Forecasting Performance}

\begin{figure}[ht]
  \centering
  \includegraphics[width=0.9\columnwidth]{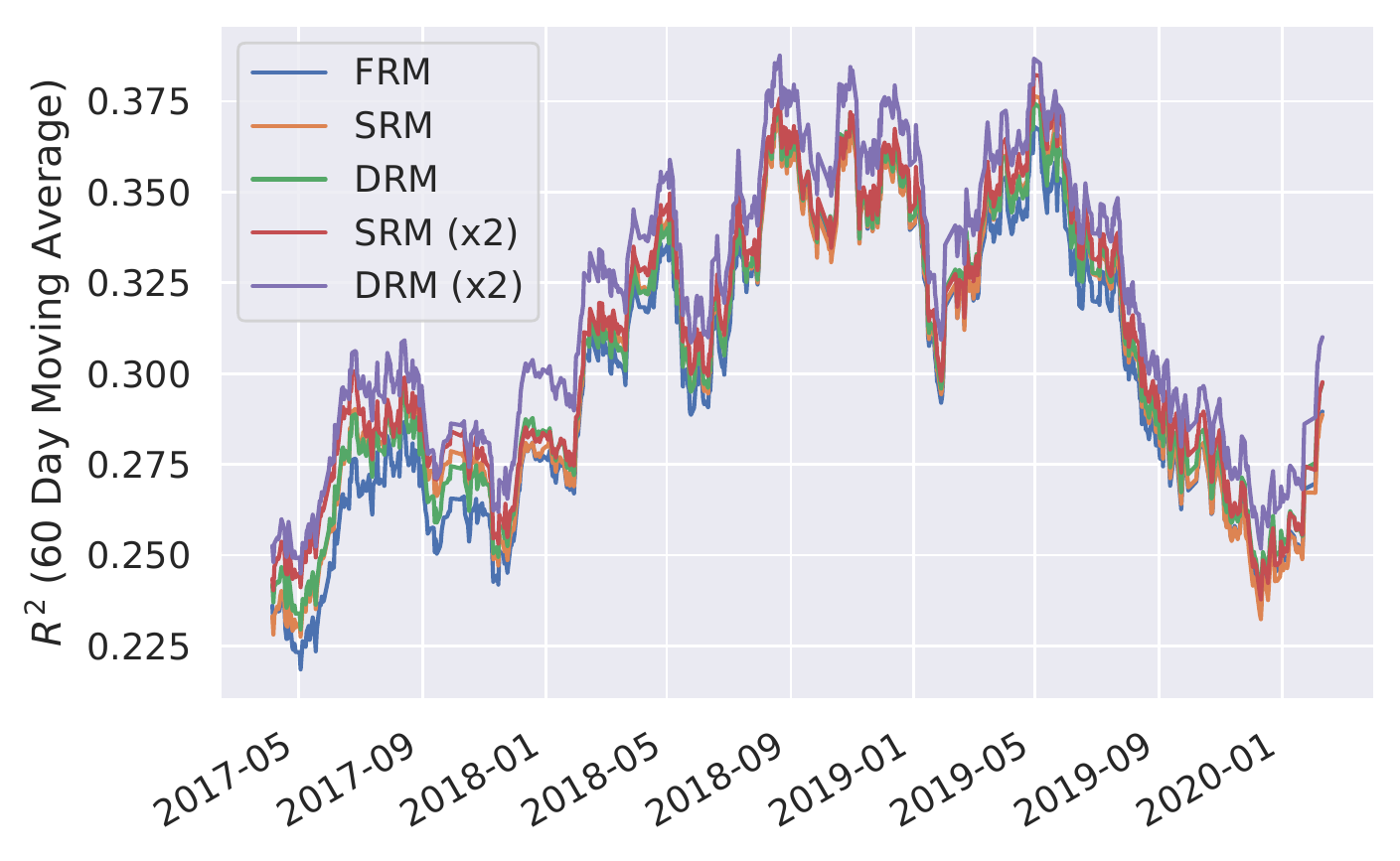}
  \caption{$R^2$ of the compared methods over the entire testing period. \textbf{DRM (x2)} consistently outperforms other methods.}
  \label{fig:r2}
\end{figure}

The risk forecasting metrics for the compared methods are summarized in Table~\ref{tab:result}. From the table, we have the following observations and conclusions:
\begin{itemize}
    \item \textbf{DRM} can improve $R^2$ by $0.6\%$ when generating the same number of risk factors compared to \textbf{FRM}, which shows the risk factors designed human experts still have room for improvement. Besides, with \textbf{DRM}'s ability to learn more risk factors, we can further improve $R^2$ by $1.9\%$. We also present $R^2$ over the entire testing period in Figure~\ref{fig:r2}, from which we can see \textbf{DRM} can consistently outperform the competitors.
    \item \textbf{DRM} can also reduce the volatility risk of both the \textbf{GMV} portfolio and the \textbf{GMV+} portfolio, by $1.0\%$ and $0.6\%$ respectively. As \textbf{GVM} portfolio is exactly determined by the covariance matrix, lower out-of-sample volatility means the estimated covariance is more accurate. This further demonstrated the learned risk factors can consistently improve covariance estimation.
    \item  The other data-driven approach for risk factor mining, \textbf{SRM}, can also improve $R^2$ but still underperforms our method. Besides, the \textbf{GMV} portfolio solved with \textbf{SRM} often performs badly. Therefore, the proposed solution is superior to \textbf{SRM} for mining latent risk factors.
\end{itemize}

\begin{table}[ht]
\caption{The performance of the compared risk models measured by explained variance of stock returns ($R^2$) and the volatility risk of two portfolios (GMV and GMV+). $\uparrow$ means the higher the better while $\downarrow$ means the opposite.}
\centering
\begin{tabular}{llll}
\toprule
Risk Model &      $R^2$ ($\uparrow$) &    GMV ($\downarrow$) &    GMV+ ($\downarrow$)\\
\midrule
Market   &       - &  18.4\% &  18.4\% \\
\hline
FRM      &  29.8\% &  12.6\% &  12.9\% \\
SRM      &  30.1\% &  18.6\% &  13.5\% \\
SRM (x2) &  30.7\% &  16.5\% &  16.2\% \\
\hline
DRM      &  30.4\% &  12.3\% &  12.9\% \\
DRM (x2) &   \textbf{31.7}\% &  \textbf{11.6}\% &  \textbf{12.3}\% \\
\bottomrule
\end{tabular}
\label{tab:result}
\end{table}

\subsubsection{Factor Characteristics}
We further present the statistical metrics of the learned risk factors in Table~\ref{tab:factor}. We can see that the average absolute t-statistics for all factors are higher than $2$ and more than $50\%$ of the time more factors have t-statistics higher than $2$,  which shows the learned risk factors are indeed significant. Further, all our risk factors have VIF scores around $1.0$, which shows there is little multicollinearity in our risk factors. Last, the auto correlation of the learned risk factors is high, which shows our method can also ensure the stability of the learned risk factors.

\begin{table}[ht]
\caption{A summary statistics of the learned risk factors in terms of significance (t-statistics), orthogonality (VIF) and stability (auto correlation).}
\centering
\begin{tabular}{lrlrr}
\toprule
Factor ID &  Mean |t| & Pct |t|>2 &    VIF &  Auto Corr. \\
\midrule
0 &     4.878 &     72.9\% &  0.984 &       0.995 \\
1 &     3.559 &     65.9\% &  1.301 &       0.995 \\
2 &     2.748 &     55.0\% &  1.180 &       0.997 \\
3 &     2.173 &     45.9\% &  1.277 &       0.998 \\
4 &     3.244 &     59.0\% &  1.309 &       0.997 \\
5 &     2.831 &     56.0\% &  1.163 &       0.966 \\
6 &     3.461 &     64.7\% &  1.007 &       0.994 \\
7 &     2.697 &     56.8\% &  1.211 &       0.966 \\
8 &     3.247 &     60.3\% &  1.293 &       0.990 \\
9 &     2.699 &     55.6\% &  1.104 &       0.985 \\
\bottomrule
\end{tabular}
\label{tab:factor}
\end{table}

\begin{figure}[ht]
  \centering
  \includegraphics[width=0.9\columnwidth]{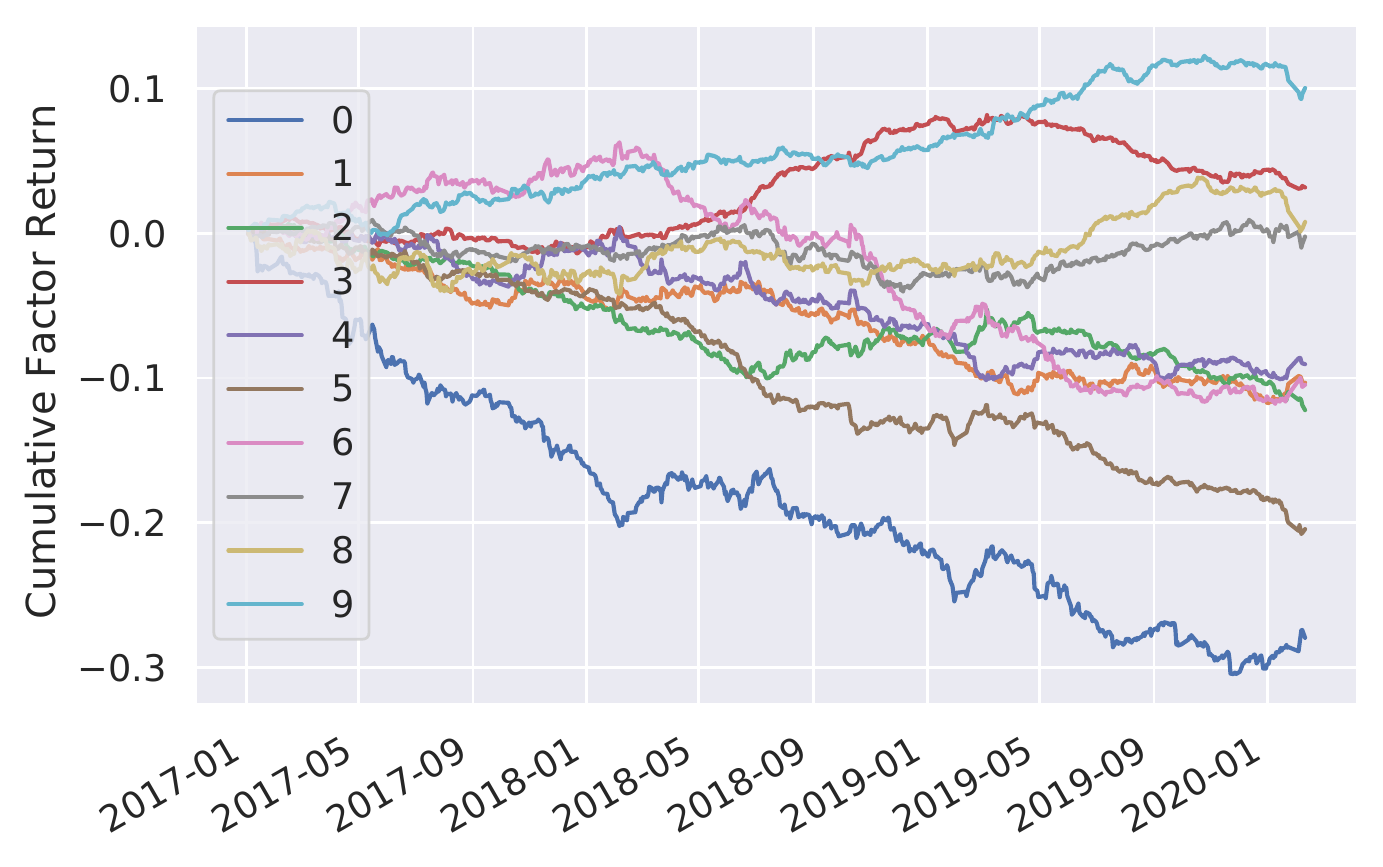}
  \caption{Cumulative factor returns (i.e., regression coefficients) of the learned risk factors.}
  \label{fig:factor_ret}
\end{figure}

Figure~\ref{fig:factor_ret} shows the cumulative factor returns, i.e., regression coefficients in Equation~\ref{eq:reg}. We can see the regression coefficients may change from positive to negative at a specific time (e.g., $3$), which indicates the investment philosophy of the entire market may have changed. This indicates the risk factors from our model do capture certain trading patterns. However, unlike fundamental risk factors, the learned risk factors still lack interpretability. We will leave the explanation of the learned risk factor as future work.

\subsection{Incremental Analysis}
In this section, we want to answer the following research questions through incremental experiments:
\begin{itemize}
    \item \textbf{RQ1}: Is GAT indeed a necessary component in the proposed neural network architecture?
    \item \textbf{RQ2}: What is the effect of the multi-collinearity regularization parameter $\lambda$ in Equation~\ref{eq:loss}?
    \item \textbf{RQ3}: Does the multi-task learning objective help stabilize the learned risk factors?
\end{itemize}

\subsubsection{RQ1} We train the proposed network with and without the GAT module in parallel. To give a more convincing conclusion, we compare these two models with different number of risk factors as learning targets. The final $R^2$ of these models are summarized in Table~\ref{tab:gat}. We can see that, introducing the GAT module can obtain higher $R^2$ in most cases. Besides, considering GAT is designed as a new transformation of the input data, we believe it will be necessary for mining more high quality risk factors if more information is introduced for risk factor mining.

\begin{table}[ht]
\caption{$R^2$ of DRM with or without GAT.}
\centering
\begin{tabular}{lllllll}
\toprule
\# Factors &     10 &     12 &     14 &     16 &     18 &     20 \\
\midrule
w/ GAT  &  \textbf{30.4}\% &  \textbf{30.6}\% &  30.8\% &  31.1\% &  \textbf{31.4}\% &  \textbf{31.7}\% \\
w/o GAT &  30.3\% &  30.5\% &  \textbf{30.9}\% &  31.1\% &  31.3\% &  31.5\% \\
\bottomrule
\end{tabular}
\label{tab:gat}
\end{table}

\subsubsection{RQ2}
\begin{figure}[ht]
  \centering
  \includegraphics[width=0.9\columnwidth]{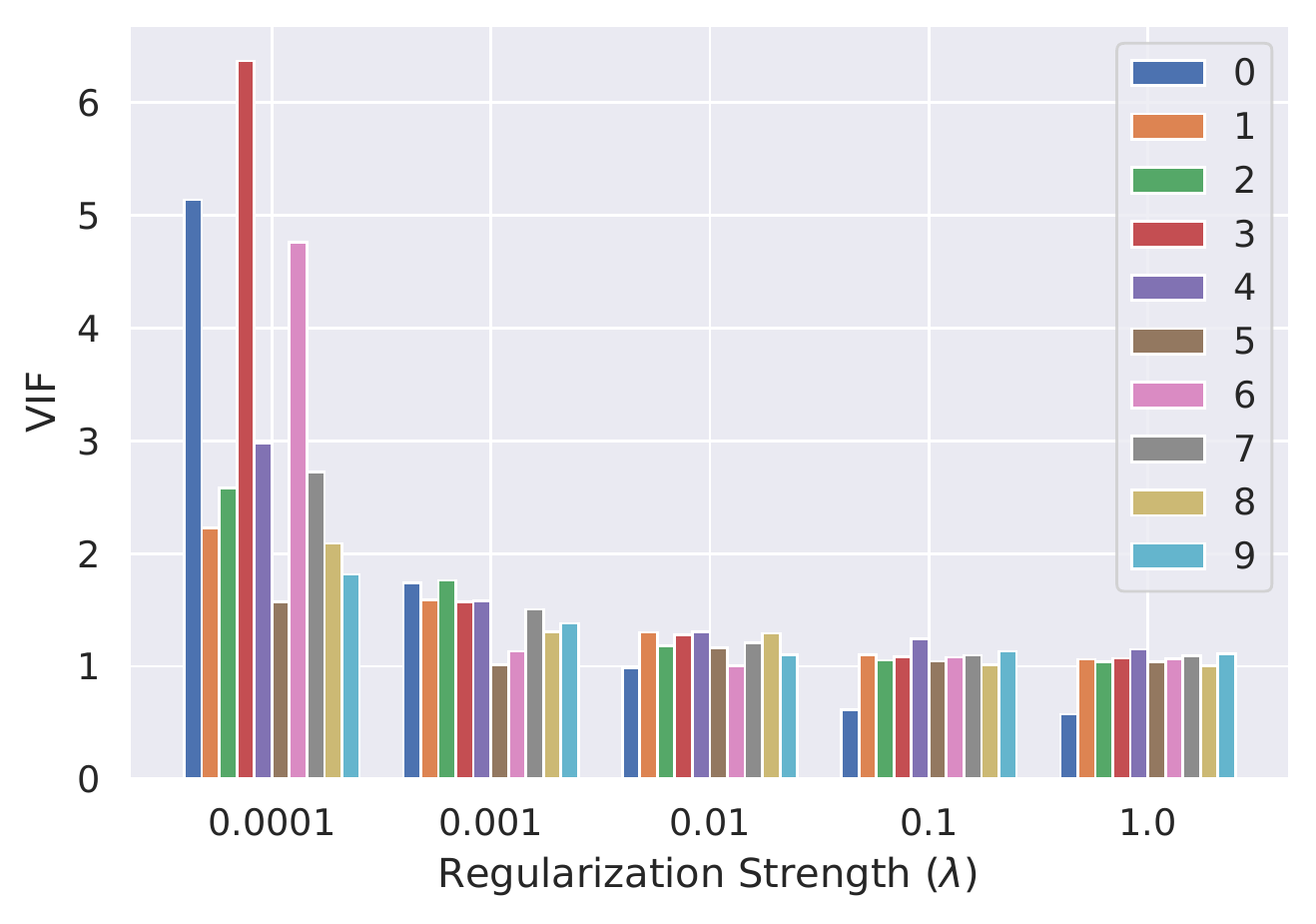}
  \caption{Variance inflation factor (VIF) of the learned risk factors with different regularization strength ($\lambda$) in Equation~\ref{eq:loss}.}
  \label{fig:vif}
\end{figure}

We further compare training the network with different regularization parameter $\lambda$. We report the VIF of the learned risk factors in Figure~\ref{fig:vif}. It can be observed that as the regularization strength increases, the VIF gradually goes down. When setting $\lambda=0.001$, the VIF can be higher than $5$, which implies the existence of high collinearity among the learned risk factors.

\subsubsection{RQ3}
\begin{figure}[ht]
  \centering
  \includegraphics[width=0.9\columnwidth]{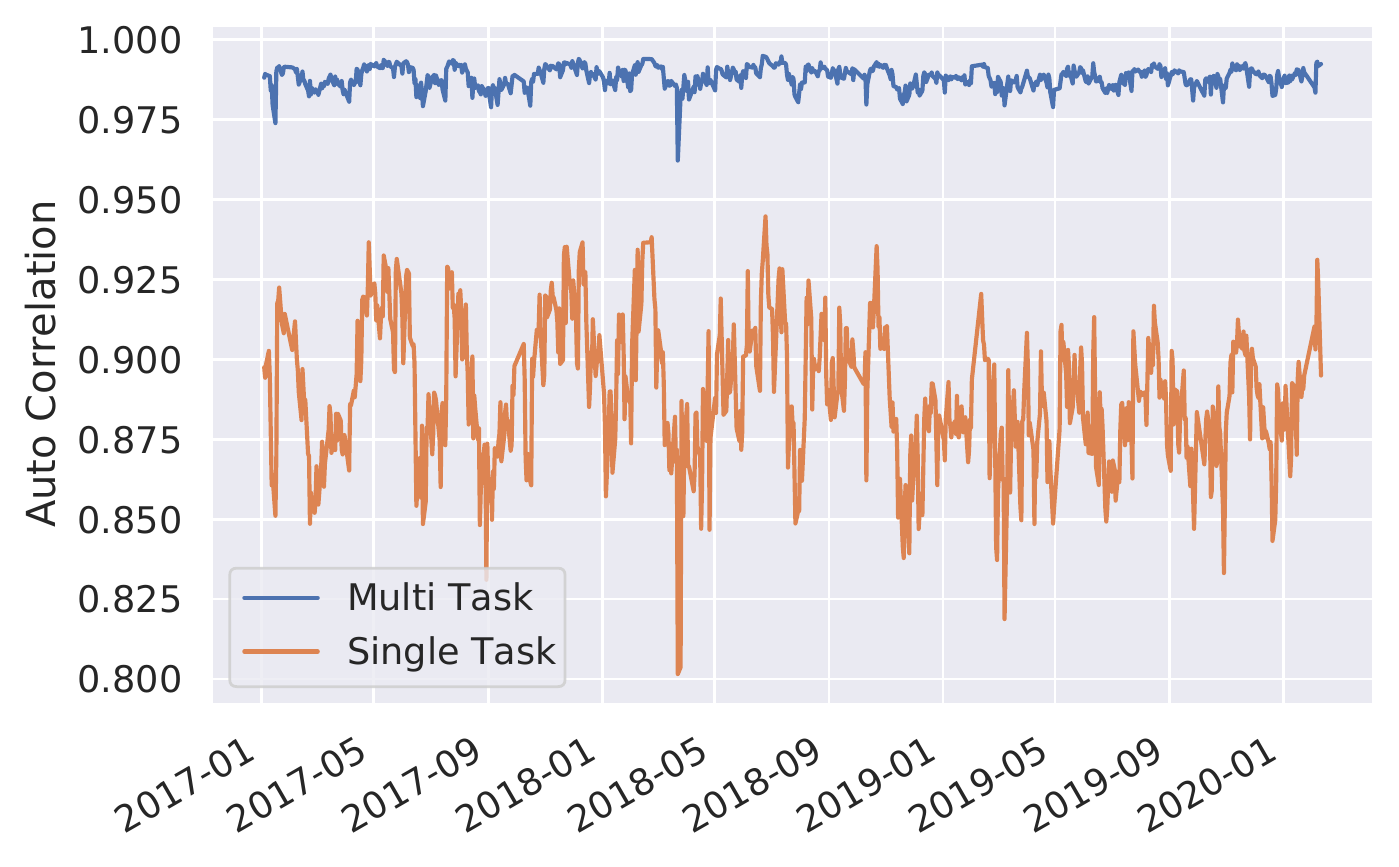}
  \caption{Auto correlation of learned risk factors with single task ($H=1$) or multi task ($H=20$) in Equation~\ref{eq:loss}.}
  \label{fig:auto_corr}
\end{figure}

Last, we verify the necessity of the multi-task objective in Equation~\ref{eq:multi-task} by comparing with the single-task objective (set $H=1$ in Equation~\ref{eq:multi-task}). Figure~\ref{fig:auto_corr} shows the average auto correlation of all learned risk factors with either single-task objective or multi-task objective. We can see that without using the multi-task objective, the risk factor becomes unstable. Therefore, we should use the multi-task objective to optimize the model.

\section{Conclusion} 
In order to improve the estimation of the covariance matrix of stock returns, we proposed a deep learning solution to effectively design risk factors. Fundamental risk model requires prodigious human professionals' effort, meanwhile statistical risk model lacks non-linear capacity and is deficient in out-of-sample performance. Our \emph{Deep Risk Model} (DRM) demonstrates impressive performance in forecasting risk with a more effective and adaptive estimate of covariance matrix. Our method also exhibits better factor characteristics in terms of high stability and low multi-collinearity. Ablation studies exemplify the necessity and superiority of our design in the architecture and learning objective.

\bibliographystyle{ACM-Reference-Format}
\bibliography{reference}

\appendix
\clearpage

\end{document}